\input mtexsis.tex %
\def\HeadLine{\hfill}
\def\FootLine{\hss\twelverm\folio\hss}
\superrefsfalse
\hsize = 17.0 truecm
\parskip = 0pt
\def\vss{\vskip 0pt plus 1fil minus 1fil}

\def\0{\hphantom{0}}
\def\naive{na\kern0.05em\"\i ve}
\def\gl{{\hbox{\raise.5ex\hbox{$>$}\kern-.8em\lower.5ex\hbox{$
<$}}}}
\def\gla{{\hbox{\raise.3ex\hbox{${\scriptscriptstyle >}$}
\kern-.5em\lower.3ex\hbox{${\scriptscriptstyle <}$}}}}
\def\vol#1{{\bf #1}}
\def\MeV{\mathop{\rm Me\kern-0.1em V}\nolimits}

\def\tr{{\rm tr}}

\overfullrule=0pt
\preprint
\thicksize = \thinsize
\pubdate{\twelvepoint May, 1993}
\pubcode{\twelvepoint IFUP-TH 17/93}
\titlepage
\title
The Three-Loop Lattice Free Energy
\endtitle
\authors
B. All\'es$^a$, M. Campostrini$^b$, A. Feo$^b$, H.
Panagopoulos$^{b,c}$
\institution{a)}{Departamento de F\'\i sica Te\'orica y del Cosmos,
Universidad de Granada, Spain}
\institution{b)}{I.N.F.N. and Dipartimento di Fisica dell'Universit\`a,
Pisa, Italy}
\institution{c)}{Dept. of Natural Sciences, Univ. of Cyprus, Nicosia,
Cyprus}

\endauthor
\abstract

We calculate the free energy of SU$(N)$ gauge theories on the lattice,
to three loops. Our result, combined with Monte Carlo data for the
average plaquette, gives a more precise estimate of the gluonic
condensate.

\endabstract
\endtitlepage

\offparens
\referencelist
\reference{Alles3}
B. All\'es, M. Campostrini, A. Feo, and H. Panagopoulos, {\it Lattice
Perturbation Theory by Computer Algebra: A Three-Loop Result for the
Topological Susceptibility}, Pisa preprint IFUP-TH 31/92
\endreference
\reference{Shifman}
M.~A.~Shifman, A.~I.~Vainshtein, and V.~I.~Zakharov, Nucl. Phys.
{\bf B147} (1979) 385, 448, 519
\endreference
\reference{DiGiacomo4}
A.~Di~Giacomo and G.~C.~Rossi, Phys.~Lett.\ {\bf B100} (1981) 481
\endreference
\reference{Amsterdam}
B.~All\'es, M.~Campostrini, A.~Feo, and H.~Panagopoulos,
Proc.\ of the {\it Lattice Gauge Theory '92\/} Conference,
Nucl.~Phys.\ B (Proc.~Suppl.) {\bf 30} (1993) 243
\endreference
\reference{Luscher2}
M.~L\"uscher and P. Weisz, Nucl. Phys. {\bf B266} (1986) 309
\endreference
\reference{Campostrini}
M.~Campostrini, G.~Curci, A.~Di~Giacomo, and G.~Paffuti,
Zeitschr. Phys. \vol{C32} (1986) 377
\endreference
\reference{DiGiacomo5}
A. Di Giacomo, H. Panagopoulos, and E. Vicari, Nucl. Phys.
{\bf B338} (1990) 294
\endreference
\reference{Campostrini3}
M. Campostrini, A. Di Giacomo, and Y. G\"und\"u\d c, Phys. Lett.
{\bf 225B} (1989) 393
\endreference
\endreferencelist

\singlespaced

In a previous work\ref{Alles3} we described a scheme for carrying out
perturbative calculations on the lattice, using a symbolic manipulation
language. We also used this scheme to obtain the first three-loop
result, related to the topological susceptibility.

In this paper, we present a three-loop calculation of the free energy
in pure SU$(N)$ gauge theories. Using this perturbative result together
with Monte Carlo data for the average plaquette we arrive at a more
precise determination of the (non- perturbative) gluon condensate
$$
G_2=-{\beta(g)\over g} \, {1\over4\pi^2 b_0} \,\sum_{a,\mu,\nu}
F_{\mu\nu}^a F_{\mu\nu}^a \,. \EQN
$$
This condensate has been widely used in phenomenological studies of
QCD, in particular in the context of the SVZ sum rules\ref{Shifman}. In
addition, the feasibility of extracting $G_2$ from a lattice simulation
is a challenging theoretical issue on its own right; the presence of
mixing with lower dimensional operators ($F_{\mu\nu} F^{\mu\nu}$ mixes
with the unit operator) represents a priori a major stumbling block. It
is important to see just to what accuracy this extraction can be pushed
given the systematic error.

The gluonic condensate is related to Monte Carlo data through the
expression
$$
\langle\Pi\rangle_{\rm MC} = Z(\beta)\,
\left(-{\pi^2\over12}\right) \,
G_2\,a^4 + \langle\Pi\rangle_{\rm pert} \,, \EQN G2
$$
where
$$
\beta = {2N\over g^2}
$$
and $\Pi$ is the plaquette operator
$$
\Pi\equiv\Pi_{x,\mu\nu}= \tr(U_{x,x{+}\mu}
U_{x{+}\mu,x{+}\mu{+}\nu} U_{x{+}\mu{+}\nu,x{+}\nu}
U_{x{+}\nu,x}). \EQN
$$
The quantities $Z(\beta)$ and $\langle\Pi\rangle_{\rm pert}$ can be
calculated perturbatively; while $Z(\beta)$ for the plaquette operator
is very close to 1 and can be left out to a rather good approximation,
$\langle\Pi\rangle_{\rm pert}$ is far from negligible, and it is
important to determine it with the best possible precision.

We write:
$$
\langle\Pi\rangle_{\rm pert} = N + {c_1\over\beta} +
{c_2\over\beta^2} + {c_3\over\beta^3} {+} \cdots \,, \EQN
$$
where $c_1= -N(N^2-1)/4$ and $c_2$ was calculated in
\Ref{DiGiacomo4}.

We have carried out the calculation of the three-loop coefficient
$c_3$. This a priori entails computing all vacuum diagrams with one
insertion of the plaquette operator; there are 63 such distinct
diagrams. The relationship between $\Pi$ and the Wilson action
$$
S=\sum_{x,\mu\nu}(N-\Pi_{x,\mu\nu}) \EQN
$$
allows a simplification of this task, by relating
$\langle\Pi\rangle_{\rm pert}$ to the partition function
\par\noindent
$Z=\int[dU]\,\exp(-\beta/2N\, S)$. We have:
$$
\eqalign{\langle\Pi\rangle_{\rm pert}&={1\over Z}
\int[dU]\,e^{-{\beta\over2N}S}\;\Pi_{x,\mu\nu} \cr
&= N-{1\over12V}\,{1\over Z}\, \int[dU]e^{-{\beta\over2N}S}\,S
= N+{2N\over12V}\,{\partial\over\partial\beta}\ln Z \,. \cr}\EQN
$$
Now, $\ln Z$ has the form
$$
\ln Z=V\left(-{3(N^2-1)\over2}\,\ln\beta + {d_1\over\beta} +
{d_2\over\beta^2}+\cdots\right) ,\EQN
$$
which implies that $c_3=-N/2\,d_2$. The calculation of $d_2$ involves
computing all 3-loop connected vacuum diagrams with no operator
insertions; there are 29 diagrams giving nonzero contributions
altogether, shown in Fig.~2.

We have computed these diagrams using our algebraic package described
in \Ref{Alles3}. The main building blocks of this package are:
\item{i)} Generating all relevant vertices.
\item{ii)} Performing contractions for each diagram and producing
the corresponding integrand. Here, as well as in i), it is
crucial to exploit all available symmetries, in order to render
the resulting expressions more compact.
\item{iii)} Integrating over loop momenta, for infinite as well
as finite lattices.

\noindent The contributions from each diagram in the Feynman gauge are
given for reference in Table I. The final result, given in Table II,
is, of course, gauge independent. Incidentally, the CPU time necessary
for the actual runs (once the development and testing stages are over)
is $\sim3$ hours on a Sun SPARC ELC to produce the integrand for some
of the more complicated diagrams and $\sim2$ hours for a highly
optimized integration on lattices up to $16^4$. (A notable exception
are diagrams with the topology of Fig.~2l, known as ``Mercedes-Benz
emblem''; the integration of these diagrams requires $\sim20$ hours
already for an $8^4$ lattice.)

A preliminary presentation of the results reported here was given in
\Ref{Amsterdam}.  Note however a discrepancy due to an error in the
preliminary version.

Using numerical results for $L^4$ lattices ($L=2,\ldots,16$) we can
extrapolate out to $L=\infty$ with a functional form: $$ r(L)=r_0 +
{r_i\over L^i} \left(+{r_j\over L^j}\right) \qquad i,j=2,3,4\EQN $$
depending on the diagram, and compare with the infinite lattice results
$r(\infty)$. (Infinite lattice results are obtained with the Schwinger
representation technique, explained in \Ref{Alles3}.)  The discrepancy
is typically a fraction of a per mille, so that $r(L)$ can be used also
for lattices with $L>16$. The actual analytic form of $r(L)$, for
$L\to\infty$, could also involve terms proportional to $(\ln
L)^n/L^m$\ref{Luscher2}; however the introduction of these terms does
not alter our extrapolations.  It should also be mentioned here that,
while lattice integrals are ultraviolet convergent, individual diagrams
may exhibit infrared divergences, which cancel only when they are
summed up. For the case at hand, the diagrams in Fig.~2o, containing a
double insertion of the one-loop renormalized gluon propagator
(Fig.~1), must be evaluated simultaneously, in order to guarantee
infrared convergence.

Having obtained the value of $c_3$ we can now reanalyze Monte Carlo
data for SU(2)\ref{Campostrini},\ref{DiGiacomo5} and
SU(3)\ref{Campostrini3}, using \Eq{G2}, to extract the gluon
condensate. We perform two series of $\chi^2$ fits, one in which $c_3$
is a parameter to be fitted, the other using the calculated value of
$c_3$ and fitting instead $c_4$. The results are listed in Table III;
in all cases we have $\chi^2/{\rm d.o.f.}\simeq 1$.  It is reassuring
to see that, despite variations in the fitted values of the parameters
$c_i$, the condensate exhibits a very stable behavior.

To conclude, our algebraic scheme has made possible this 3-loop
calculation; done by hand, such a calculation would be extremely
cumbersome, if not downright impossible. At the present stage of
development of this scheme, one can compute vacuum expectation values
of any operator to an arbitrary number of loops (modulo computer
limitations). We are currently extending the scheme to include matrix
elements with external fields; the main problem there is extracting the
analytic dependence on external momenta, which can be a rather delicate
issue beyond one loop. We will report on this subject in a future
publication.

\bigskip
\noindent
{\bf Acknowledgments. }
It is a pleasure to thank Adriano Di Giacomo for many useful
conversations. We acknowledge financial support from MURST (Italian
Ministry of the University and of Scientific and Technological
Research) and from the Spanish-Italian ``Integrated Action'' (contract
A17). B.A. also acknowledges a Spanish CICYT contract.
\vskip 2truecm
\centerline{\bf References}
\bigskip
\References
\vfill\eject
\vskip 2truecm
\singlespaced
\centerline{\bf Table I}
\smallskip
\singlespaced

The contribution to $c_3$ of individual diagrams, shown in Fig.~2, in
the Feynman gauge. We use an $L^4$ lattice and gauge group SU$(N)$.
Each entry must be multiplied by $N^5(N^2-1)\,10^{-4}$.
\bigskip
{\tenpoint
\let\tablespace=\
\ruledtable
Fig.| $L=8$ | $L=12$ | $L=16$ | $L=\infty$  \cr
a| $41.874 - 93.004/N^2 $ |%
     $41.833 - 92.971/N^2 $ |%
     $41.811 - 92.946/N^2 $ |%
     $41.775 - 92.902/N^2 $  \crnorule
 | $+ 52.045/N^4$ |%
     $+ 52.076/N^4$ |%
     $+ 52.081/N^4$ |%
     $+ 52.083/N^4$  \crnorule
b| $1.0803$ | $1.0974$ |%
     $1.1035$ | $1.1113$  \crnorule
c| $38.582 - 43.395/N^2$ |%
     $40.013 - 44.714/N^2$ |%
     $40.528 - 45.184/N^2$ |%
     $41.200 - 45.795/N^2$  \crnorule
d| $1.0806$ | $1.0975$ |%
     $1.1035$ | $1.1113$  \crnorule
e| $1.0803$ | $1.0974$ |%
     $1.1035$ | $1.1113$  \crnorule
f| $0.22019$ | $0.23398$ |%
     $0.23894$ | $0.24543$  \crnorule
g| $0.018161$ | $0.021014$ |%
     $0.022139$ | $0.023730$  \crnorule
h| $-10.464 $ |%
     $-10.696 $ |%
     $-10.781 $ |%
     $-10.892 $  \crnorule
 | $+ (1/N^2 {-} 3/N^4)\, 26.993$ |%
     $+ (1/N^2 {-} 3/N^4)\, 27.013$ |%
     $+ (1/N^2 {-} 3/N^4)\, 27.017$ |%
     $+ (1/N^2 {-} 3/N^4)\, 27.019$  \crnorule
i| $0.41123$ | $0.41825$ |%
     $0.42074$ | $0.42394$  \crnorule
j| $-1.7753$ | $-1.7876$ |%
     $-1.7880$ | $-1.7839$  \crnorule
k| $0.084525$ | $0.089997$ |%
     $0.091975$ | $0.094566$  \crnorule
l| $-0.50731$ | $-0.56679$ |%
     $-0.58988$ | $-0.62246$  \crnorule
m| $0.023925$ | $0.027060$ |%
     $0.028245$ | $0.029881$  \crnorule
n| $0.036795$ | $0.037688$ |%
     $0.038022$ | $0.038479$  \crnorule
o| $-135.184 + 290.471/N^2 $ |%
     $-136.433 + 291.877/N^2 $ |%
     $-136.861 + 292.342/N^2 $ |%
     $-137.404 + 292.919/N^2 $  \crnorule
 | $- 156.136/N^4$ |%
     $- 156.227/N^4$ |%
     $- 156.243/N^4$ |%
     $- 156.25/N^4$  \cr
Total| $-63.438 + 181.064/N^2$ |%
         $-63.517 + 181.204/N^2$ |%
         $-63.532 + 181.229/N^2$ |%
         $-63.537 + 181.240/N^2$  \crnorule
| $- 185.068/N^4$ |%
    $- 185.191/N^4$ |%
    $- 185.211/N^4$ |%
    $- 185.223/N^4$
\endruledtable
}
\bigskip\bigskip
\centerline{\bf Table II}
\smallskip

The calculated, gauge independent value of $c_3$ in SU(2) and SU(3).
\bigskip

\ruledtable
$c_3$| $L=8$ | $L=12$ | $L=16$ | $L=\infty$  \cr
$N{=}2$| $-0.28549$ | $-0.28599$ |%
           $-0.28608$ | $-0.28611$  \crnorule
$N{=}3$| $-8.8655$ | $-8.8781$ |%
           $-8.8805$ | $-8.8814$
\endruledtable
\bigskip\bigskip
\centerline{\bf Table III}
\smallskip

We list the values of $\pi^2/12N\,G_2/\Lambda_{\rm QCD}^4$ (for gauge
groups SU(2) ($\beta{=}2.45$) and SU(3) ($\beta{=}6.0$)), as
obtained from \Eq{G2} and Monte Carlo data through a series of fits, in
which: a) $c_3$ was an additional parameter to be fitted, or: b) the
exact value of $c_3$ was taken from our calculation, fitting instead
$c_4$.

\bigskip
\ruledtable
 $\pi^2/12N\,G_2/\Lambda_{\rm QCD}^4$ | SU(2) | SU(3)
\cr
a| $1.09(09)\,10^7$ | $2.79(33)\,10^8$  \crnorule
b| $1.06(10)\,10^7$ | $2.39(61)\,10^8$
\endruledtable

\bye